\documentclass{PoS}

\usepackage[utf8]{inputenc}

\title{Gamma Rays from Large-Scale Outflows in Starburst Galaxies}

\ShortTitle{Gamma Rays from Starbursts}

\author{\speaker{Gustavo E. Romero} \thanks{Member of CONICET.}\\
 Instituto Argentino de Radioastronom\'{\i}a (CONICET; CICPBA), C.C. No. 5, 1894 Villa Elisa, Argentina.\\
        E-mail: \email{romero@iar.unlp.edu.ar}}

\author{Ana L.   M\"{u}ller\\
       Instituto Argentino de Radioastronom\'{\i}a (CONICET; CICPBA), C.C. No. 5, 1894 Villa Elisa, Argentina. \\
        E-mail: \email{almuller@iar-conicet.gov.ar}}

\abstract{The combined effects of supernova explosions and stellar winds produce a hot bubble in the central
regions of starburst galaxies. As the bubble expands, it can outbreak into the galactic halo driving a
superwind that transports hot gas and fields to the intergalactic space. We present estimates of cosmic
ray generation and gamma-ray emission in both  this large-scale wind and the bow shocks created around the embedded clouds.

}

\FullConference{High Energy Phenomena in Relativistic Outflows VII - HEPRO VII\\
		9-12 July 2019\\
		Facultat de F\'isica, Universitat de Barcelona, Spain}

\begin{document}

\section{Superwinds of starburst galaxies}

Starburst galaxies undergo exceptionally high rates of star formation, often associated with galactic mergers. The enhanced star formation results in a large number of massive stars with strong winds and a high rate of supernova explosions. Both, stars and supernovae, contribute to heat up the interstellar gas which emits hard X-rays. Part of the radiation is reprocessed in the interstellar dust yielding the high infrared luminosities that are characteristic of these kind of galaxies. The energy injected by supernova explosions can additionally produce strong shock waves which accelerate cosmic rays. These particles interact with the matter, magnetic, and radiation fields producing enhanced radio and gamma-ray emission from the galactic disks (see for e.g. \cite{paglione1996,Romero2003,torres2004}).

In starburst regions, the density of core-collapse supernovae is so large as for the supernova remnants to merge. The shock fronts formed in these collisions thermalize the energy released by the stellar explosions, creating a region filled with hot gas of temperature \mbox{($T_{\rm gas}\sim 10^{8}$ K)}. The thermal energy of this matter is so high that it is unbound by the gravitational potential of the galaxy. Therefore, the hot gas expands adiabatically and escapes from the system sweeping up cooler and denser gas from the disk, driving thus a galactic superwind in which fragments of the disk are embedded \cite{chevalier1985}. 

Superwinds are complex multiphase outflows. They are formed by cool, warm, and hot gas components, as well as by dust and magnetized relativistic plasma. Currently, the existence of superwinds is supported by numerous observations. Metals are detected in the intergalactic medium, bubbles in the outflows are measured in H$_{\alpha}$,  and molecular lines plus X-ray diffuse emission consistent with gas at temperatures \mbox{$\sim 10^{7-8}$ K} are observed above the galactic disks of several starburst galaxies (see for e.g. \cite{lehnert1999,strickland2002}). Since superwinds are powered by the mechanical energy of the star forming region, their main properties scale with the star formation rate (SFR). The energy ($\dot{E}$) and mass ($\dot{M}$) injected by the superwind, consequently,  can be written as\cite{veilleux2005,tanner2017}:

\begin{equation}
 \dot{E}= \epsilon\,\dot{E}_*
\end{equation}

\begin{equation}
 \dot{M}= \dot{M}_*+\dot{M}_{\rm ISM}=\beta\,\dot{M}_*
\end{equation}

\begin{equation}
 \dot{E}_*= 7\times 10^{41} \; (\textrm{SFR}/ M_{\odot} \textrm{yr}^{-1})\;\; \textrm{erg s}^{-1}, \label{dotE}
\end{equation}

\begin{equation}
 \dot{M}_*= 0.26 \; (\textrm{SFR}/ M_{\odot} \textrm{yr}^{-1})\;\;M_{\odot} \textrm{yr}^{-1}, \label{dotM}\end{equation}

\noindent where $\epsilon$ and $\beta$ are the thermalization parameter and the mass loading factor, respectively. The thermalization parameter accounts for the degree of thermalization of the plasma in the star forming region, whereas the mass loading factor takes into account that the superwind sweeps up not only the ejecta of the stellar winds and supernova but also interstellar material.

Since the dust and gas heated up in the starburst region emit infrared radiation, it is possible to relate the SFR with the total infrared luminosity ($L_{\rm IR}$) in these galaxies \cite{kennicutt1998}:

\begin{equation}
\textrm{SFR}\approx 17 \; \frac{L_{\textrm{IR}}}{10^{11} \textrm{L}_{\odot} }\;\; M_{\odot}\; \textrm{yr}^{-1}.  \label{SFR}
\end{equation} 

Using the previous expressions, the final velocity of the superwind can be just written in terms of the thermalization and mass loading parameters \cite{veilleux2005}:

\begin{equation}
  v_{\infty}=\sqrt{\frac{2\,\epsilon\,\dot{E}_*}{\beta\,\dot{M}_*}} \approx 3000 \sqrt{\frac{\epsilon}{\beta}} \textrm{km s$^{-1}$}.
\end{equation}  

The collision of the fast superwinds with the cool swept-up halo and thick disk matter can develop large shock waves inside the superwind bubble (see Fig. 11 in \cite{strickland2002}). Radiative and slow shocks propagate forward in the direction of expansion of the bubble, whereas strong adiabatic reverse shocks propagate backwards through the superwind.

\begin{figure}
\centering
\includegraphics[trim=0cm 2cm 0cm 0cm,clip=true,width=.6\textwidth]{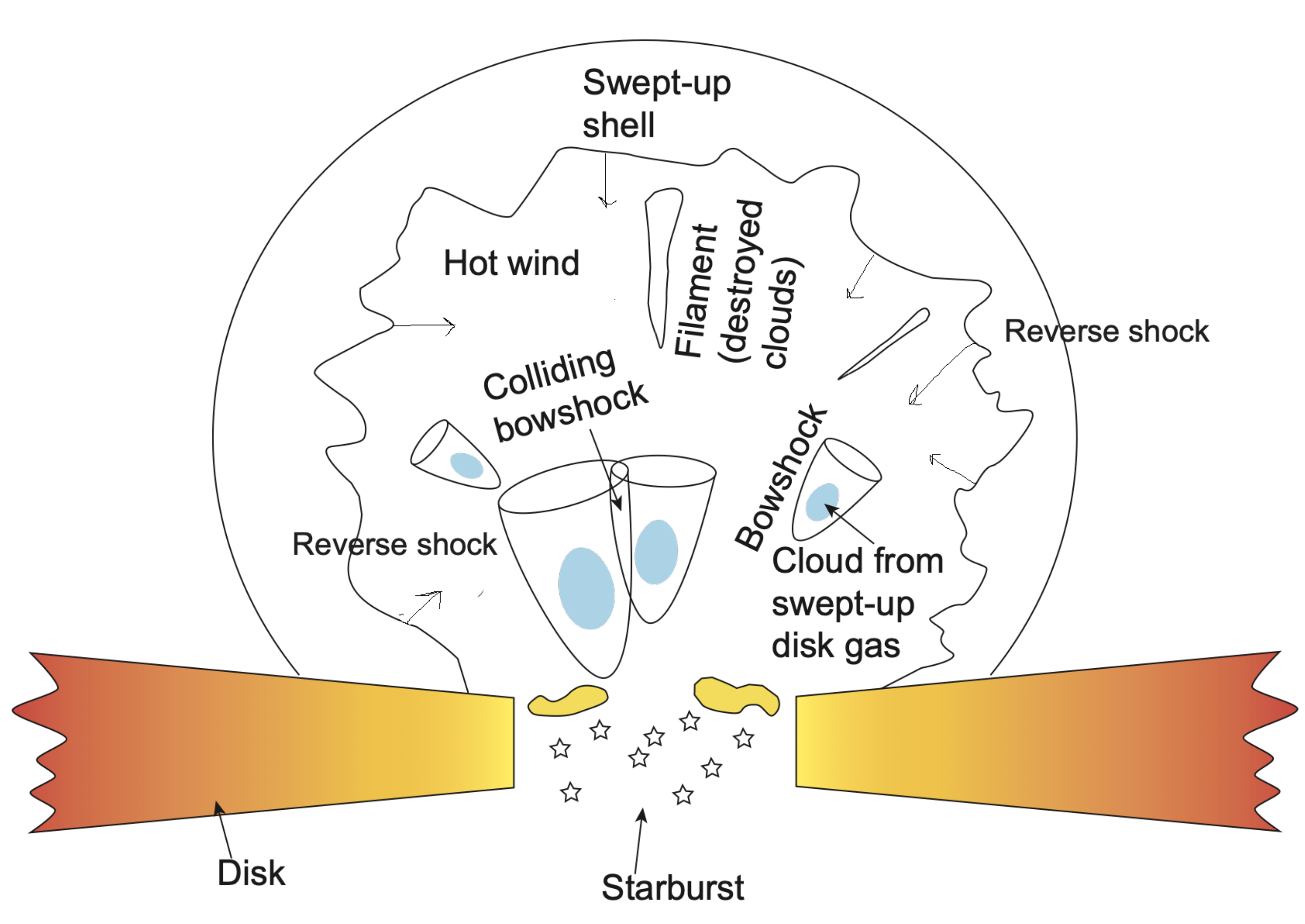}
\caption{ Sketch of superwind from the starbursts and the system of shocks associated with it. }
\label{sketch}
\end{figure}

As we mentioned before, the superwind sweeps up also matter from the disk. Several simulations show that many of these fragments remain inside the wind forming `clouds' or large `clumps' for long periods. The interaction of the supersonic wind with these clouds develops bow shocks around the latter \cite{cooper2008,cooper2009,sparre2019}. Such bow shocks have been previously studied to explain the observed thermal radiation and the emission lines detected in  halos of some starbursts (for e.g. \cite{marcolini2005}).   A sketch of the situation is presented in Figure \ref{sketch} (adapted from M\"uller et al. in preparation).

In what follows we focus on a particular starburst galaxy: the nearby southern galaxy NGC 253. This galaxy has been detected in gamma rays and has been claimed to be a source of ultra high-energy cosmic rays \cite{ackermann2012,anchordoqui1999}.

\section{NGC 253 and cosmic rays}

NGC 253 is the nearest starburst galaxy ($D \sim 3$ Mpc). It is an edge-on galaxy with a star formation rate of  \mbox{SFR $\sim 3$ M$_{\odot}$ yr$^{-1}$}. Its asymmetric superwind is detected in X-rays, H$_{\alpha}$, and CO \cite{strickland2002}. Non-thermal radio emission is observed from the galactic halo, revealing the presence of relativistic particles \cite{heesen2009}. Recently, ALMA interferometer showed that the mass loading factor of NGC 253 is very high: $\beta \sim 12$ \cite{bolatto2013}. Table 1 in \cite{romero2018} presents a summary of the physical properties of NGC 253.

Motivated by the $\gamma$-ray data collected by \textit{Fermi} satellite and H.E.S.S. in the last decade \cite{ackermann2012,lacki2011,acero2015}, we revise the model first proposed by Anchordoqui et al. (1999) \cite{anchordoqui1999}. In the scenario presented in that work, relativistic particles are accelerated at the large-scale shocks created by the interaction of the superwind with the intergalactic medium through diffusive shock acceleration (DSA) \cite{bell1978,blandford1978}. We have investigated the nature of these shocks and found that DSA is actually only possible at the strong adiabatic reverse shock (see Appendix in \cite{romero2018}). If we assume that the diffusion occurs in the Bohm's regimen, the acceleration timescale can be calculated as \cite{drury1983,romero2018}:

\begin{equation}
t_{\rm acc} \approx 2.1 \left( \frac{v_{\rm rev}}{1000\, \textrm{km s$^{-1}$}} \right)^{-2} \left( \frac{B}{\mu \textrm{G}} \right)^{-1} \left( \frac{E}{\textrm{GeV}} \right) \textrm{yr}, 
\end{equation}

\noindent where $v_{\rm rev}$ is the velocity of the reverse shock and $B$ is the strength of the magnetic field. In the case of NGC 253, the observations show a magnetic field value of \mbox{$\sim 5\,\mu$G} in the halo \cite{heesen2009}. 

Matching $t_{\rm acc}$ with the lifetime of the starburst ($\sim 10$ Myr), the maximum energies for accelerated hadrons are found: \mbox{$E_{\rm max} \sim 10^{16}$ eV} for protons and \mbox{$\sim 4\times10^{17}$ eV} for iron nuclei. These energies are smaller than those claimed by \cite{anchordoqui1999,anchordoqui2018}, which are based on overestimates of the shock velocity and the magnetic field, respectively. We have then obtained the particle distributions and built the spectral energy distributions (SEDs) of the associated radiation taking into the account the observational constraints given by \cite{lacki2011}. 

\section{Spectral energy distributions}

\begin{figure}
\centering
\includegraphics[trim=0cm 0cm 0cm 0cm,clip=true,width=.6\textwidth]{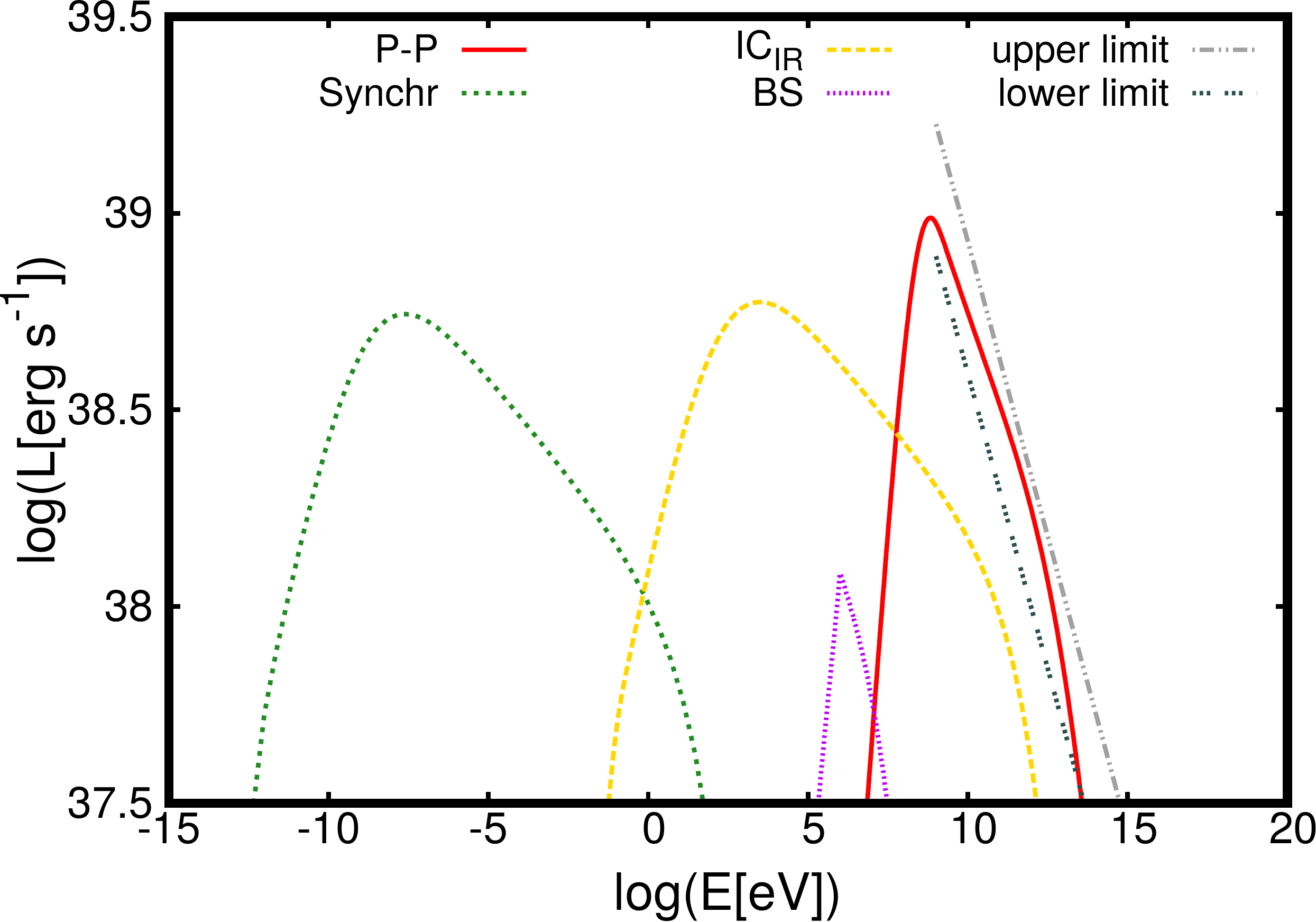}
\caption{ SED for non-thermal emission from the reverse shock starburst superwinds produced by synchrotron, inverse Compton and proton-proton inelastic collisions. The dashed lines indicate the upper and lower limits imposed by \emph{Fermi} satellite and HESS observatory; see ref. \cite{romero2018}.}
\label{fig:SED1}
\end{figure}

The resulting SEDs  are shown in Fig.  \ref{fig:SED1} \cite{romero2018}. We conclude that a significant (of about $\sim 50\%$) part of the total  $\gamma$-ray radiation observed from this source can be explained by proton-proton inelastic collisions in the superwind. 

\begin{figure}
\centering
\includegraphics[trim=0cm 0cm 0cm 0cm,clip=true,width=.6\textwidth]{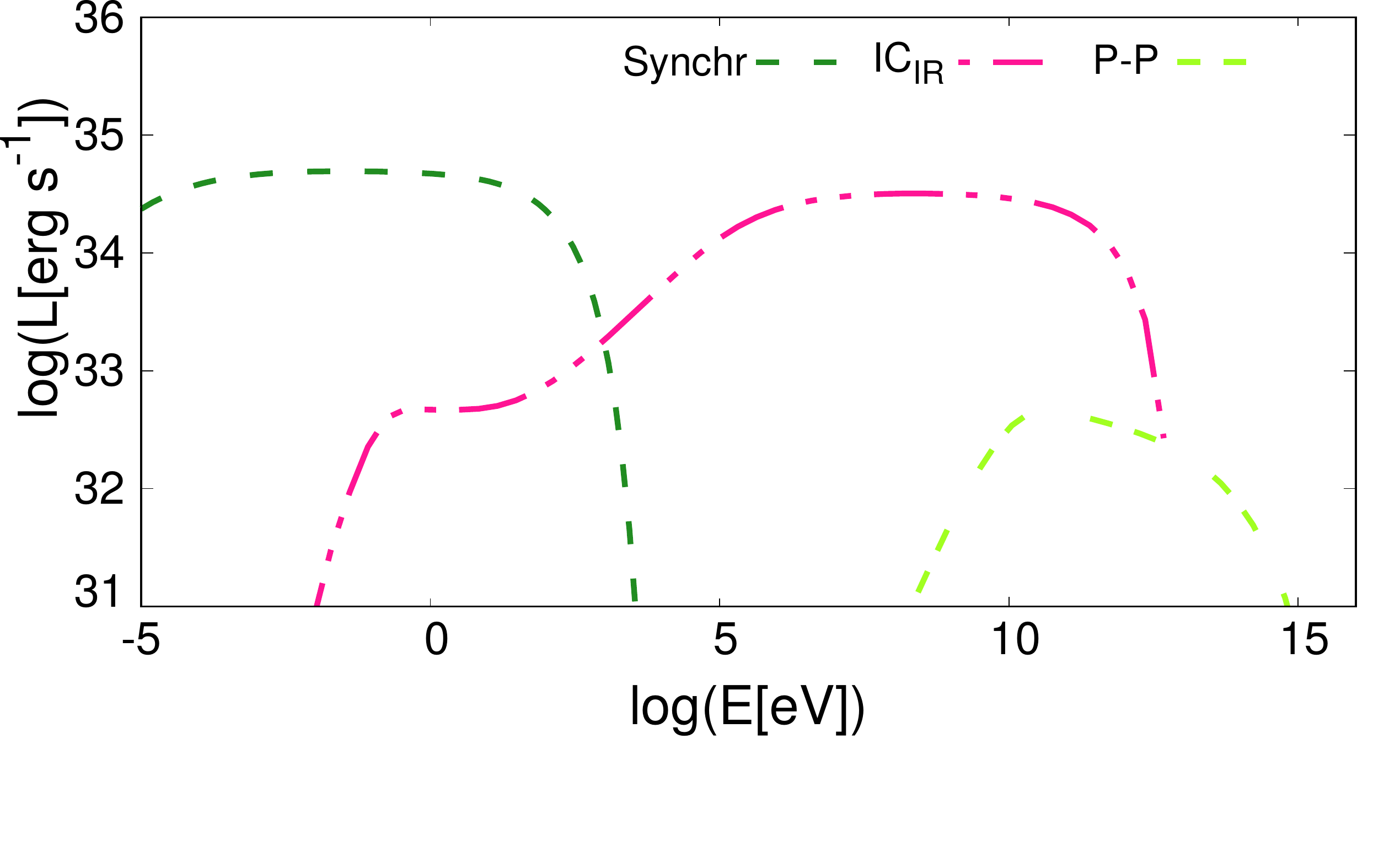}
\caption{Preliminary SED for non-thermal emission from a bow shock in the superwind of the starburst NGC 253. The radiation is produced by synchrotron, inverse Compton, and proton-proton inelastic collisions.}
\label{fig:SED2}
\end{figure}

On the other hand, we have also studied the possibility of accelerating particles at the bow shocks around the fragments of disk inside the superwind (M\"uller et al. in preparation). The maximum energies reached by the particles in this scenario are smaller than those obtained for the large-scale shocks in the superwind bubble. We explore the parameter space generating with several different models. As a result we found that the $\gamma$-ray luminosity is high enough to allow  the detection of some bow shocks as point-like sources with the forthcoming Cherenkov Telescope Array (see Fig. \ref{fig:SED2}). We also found that these bows shocks can be point-like X-ray sources for current X-ray instruments. Several of such sources have been actually observed in the halo of both NGC 253 and the similar galaxy M82 (e.g.  ref. \cite{wik2014}). 

\section{Conclusions}

We can conclude that:

\begin{itemize}
\item NGC 253 and other similar starburst galaxies are sources of cosmic rays below the ankle of the cosmic ray spectrum. If cosmic rays of higher energies are produced in these objects, as suggested by recent results indicating a high-metallicity content towards the end of the cosmic ray spectrum and some positional correlations, then such particles should be accelerated not at the superwind, but in the inner source (putative black hole at the center of the galaxy) or by gamma-ray bursts/magnetars in the disk.

\item We estimate that $\gamma$ rays up to $\sim10^{15}$ eV are accelerated inside the halos of starburst galaxies. This emission can be as important as the emission from the disk, but reaching higher energies

\item Discrete $\gamma$-ray sources should exist in the halo of starbursts because of the interaction of the superwind with fragments of disk. Some X-ray sources in the halo might be associated with such phenomenon as well.
\end{itemize}

All this shows that starburst galaxies are complex very high-energy sources that deserve thorough investigation in the light of multi-wavelength observations in order to clarify the true nature of the various physical processes occurring in them.  

\section*{Acknowledgements}This work was supported by the Argentine agency CONICET (PIP 2014-00338)
and the Spanish Ministerio de Econm\'ia y Competitividad
(MINECO/FEDER, UE) under grant AYA2016-76012-C3-1-P.

\bibliographystyle{JHEP}
\bibliography{references}

\providecommand{\href}[2]{#2}\begingroup\raggedright\begin{thebibliography}{10}

\bibitem{paglione1996}
T.~A.~D. Paglione, A.~P. Marscher, J.~M. Jackson and D.~L. Bertsch,
  \emph{{Diffuse Gamma-Ray Emission from the Starburst Galaxy NGC 253}},
  \href{https://doi.org/10.1086/176969}{\emph{Astrophys. J.} {\bfseries 460}
  (1996) 295}.

\bibitem{Romero2003}
G.~E. Romero, D.~F. Torres, M.~M.~K. Bernado and I.~F. Mirabel, \emph{{Hadronic
  gamma-ray emission from windy microquasars}},
  \href{https://doi.org/10.1051/0004-6361:20031314-1}{\emph{Astron. Astrophys.}
  {\bfseries 410} (2003) L1}
  [\href{https://arxiv.org/abs/astro-ph/0309123}{{\ttfamily
  astro-ph/0309123}}].

\bibitem{torres2004}
D.~F. Torres, E.~Domingo-Santamaria and G.~E. Romero, \emph{{High-energy
  gamma-rays from stellar associations}},
  \href{https://doi.org/10.1086/381803}{\emph{Astrophys. J.} {\bfseries 601}
  (2004) L75} [\href{https://arxiv.org/abs/astro-ph/0312128}{{\ttfamily
  astro-ph/0312128}}].

\bibitem{chevalier1985}
R.~A. Chevalier and A.~W. Clegg, \emph{{Wind from a starburst galaxy nucleus}},
  \href{https://doi.org/10.1038/317044a0}{\emph{Nature} {\bfseries 317} (1985)
  44}.

\bibitem{lehnert1999}
M.~D. Lehnert, T.~M. Heckman and K.~A. Weaver, \emph{{Very extended x-ray and
  h-alpha emission in m82: implications for the superwind phenomenon}},
  \href{https://doi.org/10.1086/307762}{\emph{Astrophys. J.} {\bfseries 523}
  (1999) 575} [\href{https://arxiv.org/abs/astro-ph/9904227}{{\ttfamily
  astro-ph/9904227}}].

\bibitem{strickland2002}
D.~K. Strickland, T.~M. Heckman, K.~A. Weaver, C.~G. Hoopes and M.~Dahlem,
  \emph{{Chandra observations of ngc 253. 2. On the origin of diffuse x-ray
  emission in the halos of starburst galaxies}},
  \href{https://doi.org/10.1086/338889}{\emph{Astrophys. J.} {\bfseries 568}
  (2002) 689} [\href{https://arxiv.org/abs/astro-ph/0111511}{{\ttfamily
  astro-ph/0111511}}].

\bibitem{veilleux2005}
S.~Veilleux, G.~Cecil and J.~Bland-Hawthorn, \emph{{Galactic winds}},
  \href{https://doi.org/10.1146/annurev.astro.43.072103.150610}{\emph{Ann. Rev.
  Astron. Astrophys.} {\bfseries 43} (2005) 769}
  [\href{https://arxiv.org/abs/astro-ph/0504435}{{\ttfamily
  astro-ph/0504435}}].

\bibitem{tanner2017}
R.~Tanner, G.~Cecil and F.~Heitsch, \emph{Scaling relations of starburst-driven
  galactic winds},
  \href{https://doi.org/10.3847/1538-4357/aa78a8}{\emph{Astrophys. J.}
  {\bfseries 843} (2017) 137}.

\bibitem{kennicutt1998}
R.~C. Kennicutt, Jr., \emph{{Star formation in galaxies along the Hubble
  sequence}}, \href{https://doi.org/10.1146/annurev.astro.36.1.189}{\emph{Ann.
  Rev. Astron. Astrophys.} {\bfseries 36} (1998) 189}
  [\href{https://arxiv.org/abs/astro-ph/9807187}{{\ttfamily
  astro-ph/9807187}}].

\bibitem{cooper2008}
J.~L. Cooper, G.~V. Bicknell, R.~S. Sutherland and J.~Bland-Hawthorn,
  \emph{{Three-Dimensional Simulations of a Starburst-Driven Galactic Wind}},
  \href{https://doi.org/10.1007/s10509-007-9526-4}{\emph{Astrophys. J.} (2007)
  } [\href{https://arxiv.org/abs/0710.5437}{{\ttfamily 0710.5437}}].

\bibitem{cooper2009}
J.~L. Cooper, G.~V. Bicknell, R.~S. Sutherland and J.~Bland-Hawthorn,
  \emph{Starburst-driven galactic winds: Filament formation and emission
  processes},
  \href{https://doi.org/10.1088/0004-637x/703/1/330}{\emph{Astrophys. J.}
  {\bfseries 703} (2009) 330}.

\bibitem{sparre2019}
M.~Sparre, C.~Pfrommer and M.~Vogelsberger, \emph{{The physics of multiphase
  gas flows: fragmentation of a radiatively cooling gas cloud in a hot wind}},
  \href{https://doi.org/10.1093/mnras/sty3063}{\emph{Mon. Not. Roy. Astron.
  Soc.} {\bfseries 482} (2019) 5401}
  [\href{https://arxiv.org/abs/1807.07971}{{\ttfamily 1807.07971}}].

\bibitem{marcolini2005}
A.~Marcolini, D.~K. Strickland, A.~D'Ercole, T.~M. Heckman and C.~G. Hoopes,
  \emph{{The Dynamics and high-energy emission of conductive gas clouds in
  supernova-driven galactic superwinds}},
  \href{https://doi.org/10.1111/j.1365-2966.2005.09343.x}{\emph{Mon. Not. Roy.
  Astron. Soc.} {\bfseries 362} (2005) 626}
  [\href{https://arxiv.org/abs/astro-ph/0506645}{{\ttfamily
  astro-ph/0506645}}].

\bibitem{ackermann2012}
M.~Ackermann, M.~Ajello, A.~Allafort, L.~Baldini, J.~Ballet, D.~Bastieri
  et~al., \emph{Gev observations of star-forming galaxies with the
  \textit{Fermi} large area telescope},
  \href{https://doi.org/10.1088/0004-637x/755/2/164}{\emph{Astrophys. J.}
  {\bfseries 755} (2012) 164}.

\bibitem{anchordoqui1999}
L.~A. Anchordoqui, G.~E. Romero and J.~A. Combi, \emph{{Heavy nuclei at the end
  of the cosmic ray spectrum?}},
  \href{https://doi.org/10.1103/PhysRevD.60.103001}{\emph{Phys. Rev.}
  {\bfseries D60} (1999) 103001}
  [\href{https://arxiv.org/abs/astro-ph/9903145}{{\ttfamily
  astro-ph/9903145}}].

\bibitem{heesen2009}
V.~Heesen, R.~Beck, M.~Krause and R.~J. Dettmar, \emph{{Cosmic rays and the
  magnetic field in the nearby starburst galaxy NGC 253 I. The distribution and
  transport of cosmic rays}},
  \href{https://doi.org/10.1051/0004-6361:200810543}{\emph{Astron. Astrophys.}
  {\bfseries 494} (2009) 563}
  [\href{https://arxiv.org/abs/0812.0346}{{\ttfamily 0812.0346}}].

\bibitem{bolatto2013}
A.~D. Bolatto et~al., \emph{{The Starburst-Driven Molecular Wind in NGC 253 and
  the Suppression of Star Formation}},
  \href{https://doi.org/10.1038/nature12351}{\emph{Nature} {\bfseries 499}
  (2013) 450} [\href{https://arxiv.org/abs/1307.6259}{{\ttfamily 1307.6259}}].

\bibitem{romero2018}
G.~E. Romero, A.~L. Müller and M.~Roth, \emph{{Particle acceleration in the
  superwinds of starburst galaxies}},
  \href{https://doi.org/10.1051/0004-6361/201832666}{\emph{Astron. Astrophys.}
  {\bfseries 616} (2018) A57}
  [\href{https://arxiv.org/abs/1801.06483}{{\ttfamily 1801.06483}}].

\bibitem{lacki2011}
B.~C. Lacki, T.~A. Thompson, E.~Quataert, A.~Loeb and E.~Waxman, \emph{{On The
  GeV \& TeV Detections of the Starburst Galaxies M82 \& NGC 253}},
  \href{https://doi.org/10.1088/0004-637X/734/2/107}{\emph{Astrophys. J.}
  {\bfseries 734} (2011) 107}
  [\href{https://arxiv.org/abs/1003.3257}{{\ttfamily 1003.3257}}].

\bibitem{acero2015}
{\scshape Fermi-LAT} collaboration, \emph{{Fermi Large Area Telescope Third
  Source Catalog}},
  \href{https://doi.org/10.1088/0067-0049/218/2/23}{\emph{Astrophys. J. Suppl.}
  {\bfseries 218} (2015) 23}
  [\href{https://arxiv.org/abs/1501.02003}{{\ttfamily 1501.02003}}].

\bibitem{bell1978}
A.~R. Bell, \emph{{The Acceleration of cosmic rays in shock fronts. I}},
  {\emph{Mon. Not. Roy. Astron. Soc.} {\bfseries 182} (1978) 147}.

\bibitem{blandford1978}
R.~D. Blandford and J.~P. Ostriker, \emph{{Particle acceleration by
  astrophysical shocks}},
  \href{https://doi.org/10.1086/182658}{\emph{Astrophys. J. L.} {\bfseries 221}
  (1978) L29}.

\bibitem{drury1983}
L.~O. Drury, \emph{{An introduction to the theory of diffusive shock
  acceleration of energetic particles in tenuous plasmas}},
  \href{https://doi.org/10.1088/0034-4885/46/8/002}{\emph{Rept. Prog. Phys.}
  {\bfseries 46} (1983) 973}.

\bibitem{anchordoqui2018}
L.~A. Anchordoqui, \emph{{Acceleration of ultrahigh-energy cosmic rays in
  starburst superwinds}},
  \href{https://doi.org/10.1103/PhysRevD.97.063010}{\emph{Phys. Rev.}
  {\bfseries D97} (2018) 063010}
  [\href{https://arxiv.org/abs/1801.07170}{{\ttfamily 1801.07170}}].

\bibitem{wik2014}
D.~R. {Wik}, B.~D. {Lehmer}, A.~E. {Hornschemeier}, M.~{Yukita}, A.~{Ptak},
  A.~{Zezas} et~al., \emph{{Spatially Resolving a Starburst Galaxy at Hard
  X-Ray Energies: NuSTAR, Chandra, and VLBA Observations of NGC 253}},
  \href{https://doi.org/10.1088/0004-637X/797/2/79}{\emph{Astrophys. J}
  {\bfseries 797} (2014) 79} [\href{https://arxiv.org/abs/1411.1089}{{\ttfamily
  1411.1089}}].

\end{thebibliography}\endgroup

\end{document}